\documentclass[runningheads,a4paper]{llncs}

\usepackage{amssymb}
\usepackage{graphicx}
\graphicspath{{figures/}}
\DeclareGraphicsExtensions{.jpg}
\usepackage{caption}
\usepackage{array}
\usepackage{stfloats}
\usepackage{multirow}
\usepackage[export]{adjustbox}

\usepackage{caption,subfig}

\usepackage{cite}

\usepackage{url}
\urldef{\mailsa}\path|{alfred.hofmann, ursula.barth, ingrid.haas, frank.holzwarth,|
\urldef{\mailsb}\path|anna.kramer, leonie.kunz, christine.reiss, nicole.sator,|
\urldef{\mailsc}\path|erika.siebert-cole, peter.strasser, lncs}@springer.com|    
\newcommand{\keywords}[1]{\par\addvspace\baselineskip
\noindent\keywordname\enspace\ignorespaces#1}

\begin{document}

\mainmatter  

\title{How Data Volume Affects \\Spark Based Data Analytics on a Scale-up Server}


%
%
\author{Ahsan Javed Awan$^1$\and Mats Brorsson$^1$\and Vladimir Vlassov$^1$\and Eduard Ayguade$^2$}
%

\institute{
$^1$KTH Royal Institute of Technology,\\
Software and Computer Systems Department(SCS),\\
\email{\{ajawan,matsbror,vladv\}@kth.se}\\
$^2$Technical University of Catalunya (UPC),\\
Computer Architecture Department,\\
\email{eduard@ac.upc.edu}\\
}

%
%

\toctitle{Lecture Notes in Computer Science}
\tocauthor{Authors' Instructions}
\maketitle

\begin{abstract}
\emph{Sheer increase in volume of data over the last decade has triggered research in cluster computing frameworks that enable web enterprises to extract big insights from big data. While Apache Spark is gaining popularity for exhibiting superior scale-out performance on the commodity machines, the impact of data volume on the performance of Spark based data analytics in scale-up configuration is not well understood. We present a deep-dive analysis of Spark based applications on a large scale-up server machine. Our analysis reveals that Spark based data analytics are DRAM bound and do not benefit by using more than 12 cores for an executor. By enlarging input data size, application performance degrades significantly due to substantial increase in wait time during I/O operations and garbage collection, despite 10\% better instruction retirement rate (due to lower L1 cache misses and higher core utilization). We match memory behaviour with the garbage collector to improve performance of applications between 1.6x to 3x.}
\end{abstract}

\section{Introduction}
With a deluge in the volume and variety of data collected, large-scale web enterprises (such as Yahoo, Facebook, and Google) run big data analytics applications using clusters of commodity servers. 
However, it has been recently reported that using clusters is a case of over-provisioning since a majority of analytics jobs do not process huge data sets and that modern scale-up servers are adequate to run analytics jobs~\cite{Scale_up_vs_Scale_out_for_Hadoop}. Additionally, commonly used predictive analytics such as  machine learning algorithms work on filtered datasets that easily fit into memory of modern scale-up servers. Moreover the today's scale-up servers can have CPU, memory and persistent  storage resources in abundance at affordable prices. Thus we envision small cluster of scale-up servers to be the preferable choice of enterprises in near future. 

While Phoenix~\cite{Phoenix_Rebirth}, Ostrich~\cite{Tiled_mr} and Polymer~\cite{Polymer} are specifically designed to exploit the potential of a single scale-up server, they don't scale-out to multiple scale-up servers. Apache Spark~\cite{Spark} is getting popular in industry because it enables in-memory processing, scales out to large number of commodity machines and provides a unified framework for batch and stream processing of big data workloads. However it's performance on modern scale-up servers is not fully  understood. A recent study~\cite{Awan} characterizes the performance of Spark based data analytics on a scale-up server but it does not quantify the impact of data volume. Knowing the limitations of modern scale-up servers for Spark based data analytics will help in achieving the future goal of improving the performance of Spark based data analytics on small clusters of scale-up servers. In this paper, we answer the following questions concerning Spark based data analytics running on modern scale-up servers:

\begin{itemize}
\item Do Spark based data analytics benefit from using larger scale-up servers?
\item How severe is the impact of garbage collection on performance of Spark based data analytics?
\item Is file I/O detrimental to Spark based data analytics performance? 
\item How does data size affect the micro-architecture performance of Spark based data analytics?
\end{itemize}

To answer the above questions, we use empirical evaluation of Apache Spark based benchmark applications on a modern scale-up server. Our contributions are:

\begin{itemize}

\item We evaluate the impact of data volume on the performance of Spark based data analytics running on a scale-up server. 
\item We find the limitations of using Spark on a scale-up server with large volumes of data.
\item We quantify the variations in micro-architectural performance of applications across different data volumes. 
\end{itemize}

\section{\textbf{Background}}

Spark is a cluster computing framework that uses Resilient Distributed Datasets (RDDs)~\cite{Spark} which are immutable collections of objects spread across a cluster. Spark programming model is based on higher-order functions that execute user-defined functions in parallel. These higher-order functions are of two types: Transformations and Actions. Transformations are lazy operators that create new RDDs. Actions launch a computation on RDDs and generate an output. When a user runs an action on an RDD, Spark first builds a DAG of stages from the RDD lineage graph. Next, it splits the DAG into stages that contain pipelined transformations with narrow dependencies. Further, it divides each stage into tasks. A task is a combination of data and computation. Tasks are assigned to executor pool threads. Spark executes all tasks within a stage before moving on to the next stage.

Spark runs as a Java process on a Java Virtual Machine(JVM). The JVM has a heap space which is divided into young and old generations.  
The young generation keeps short-lived objects while the old generation holds objects with longer lifetimes. The young generation is further divided into eden, survivor1 and survivor2 spaces. When the eden space is full, a minor garbage collection (GC) is run on the eden space and objects that are alive from eden and survivor1 are copied to survivor2. The survivor regions are then swapped. If an object is old enough or survivor2 is full, it is moved to the old space. Finally when the old space is close to full, a full GC operation is invoked.

\section{\textbf{Methodology}}
\subsection{\textbf{Benchmarks}}

Table \ref{Benchmarks} shows the list of benchmarks along with transformations and actions involved. We used Spark versions of the following benchmarks from BigDataBench~\cite{BigDataBench}. Big Data Generator Suite (BDGS), an open source tool was used to generate synthetic datasets based on raw data sets~\cite{BDGS}. 

\begin{itemize}
\item \textbf{Word Count (Wc)} counts the number of occurrences of each word in a text file. The input is unstructured Wikipedia Entries.
\item \textbf{Grep (Gp)} searches for the keyword ``The'' in a text file and filters out the lines with matching strings to the output file.  It works on unstructured Wikipedia Entries.
\item \textbf{Sort (So)} ranks records by their key. Its input is a set of samples. Each sample is represented as a numerical d-dimensional vector.

\item \textbf{Naive Bayes (Nb)} uses semi-structured Amazon Movie Reviews data-sets for sentiment classification. We use only the classification part of the benchmark in our experiments.

\textbf{K-Means (Km)} clusters data points into a predefined number of clusters. We run the benchmark for 4 iterations with 8 desired clusters. Its input is structured records, each represented as a numerical d-dimensional vector.

\end{itemize}

\begin{table}[!ht]
\renewcommand{\arraystretch}{1.3}
\caption{Benchmarks.}
\label{Benchmarks}
\centering
\begin{tabular}{p{2cm}p{2cm}|p{3cm}|p{3cm}}
\hline
\multicolumn{2}{l|}{\textbf{Benchmarks}} & \textbf{Transformations} & \textbf{Actions} \\ \hline
\multicolumn{1}{l|}{Micro-benchmarks} & Word count & map, reduceByKey & saveAsTextFile \\\cline{2-4}
\multicolumn{1}{l|}{} & Grep & filter & saveAsTextFile \\ \cline{2-4} 
\multicolumn{1}{l|}{} & Sort & map, sortByKey & saveAsTextFile \\ \hline
\multicolumn{1}{l|}{Classification} & Naive Bayes & map & collect \\
\multicolumn{1}{l|}{} &  &  & saveAsTextFile \\ \hline
\multicolumn{1}{l|}{Clustering} & K-Means & map, filter & takeSample \\
\multicolumn{1}{l|}{} &  & mapPartitions & collectAsMap \\
\multicolumn{1}{l|}{} &  & reduceByKey & collect \\ \hline
\end{tabular}
\end{table}

\subsection{\textbf{System Configuration}}

Table~\ref{hardware} shows details about our test machine. Hyper-Threading and Turbo-boost are disabled through BIOS because it is difficult to interpret the micro-architectural data with these features enabled~\cite{HT_disabled}. With Hyper-Threading and Turbo-boost disabled, there are 24 cores in the system operating at the frequency of 2.7 GHz.

\begin{table}[!ht]
\renewcommand{\arraystretch}{1.3}
\caption{Machine Details.}
\label{hardware}
\centering
\begin{tabular}{l|l|p{6.5cm}}
\hline
\textbf{Component} & \multicolumn{2}{c}{\textbf{Details}} \\ \hline
Processor & \multicolumn{2}{l}{Intel Xeon E5-2697 V2, Ivy Bridge micro-architecture} \\ \hline
\multirow{6}{*}{} & Cores & 12 @ 2.7 GHz (Turbo upto 3.5 GHz) \\ \cline{2-3} 
 & Threads & 2 per core \\ \cline{2-3} 
 & Sockets & 2 \\ \cline{2-3} 
 & L1 Cache & 32 KB for instructions and 32 KB for data per core \\ \cline{2-3} 
 & L2 Cache & 256 KB per core \\ \cline{2-3} 
 & L3 Cache (LLC) & 30 MB per socket \\ \hline
Memory & \multicolumn{2}{l}{2 x 32 GB, 4 DDR3 channels, Max BW 60 GB/s} \\ \hline
OS & \multicolumn{2}{l}{Linux kernel version 2.6.32} \\ \hline
JVM & \multicolumn{2}{l}{Oracle Hotspot JDK version 7u71} \\ \hline
Spark & \multicolumn{2}{l}{Version 1.3.0} \\ \hline
\end{tabular}
\end{table}

Table~\ref{parameters} also lists the parameters of JVM and Spark. For our experiments, we use HotSpot JDK version 7u71 configured in server mode (64 bit). The Hotspot JDK provides several parallel/concurrent GCs out of which we use three combinations: (1) Parallel Scavenge (PS) and Parallel Mark Sweep; (2) Parallel New and Concurrent Mark Sweep; and  (3) G1 young and G1 mixed for young and old generations respectively. The details on each algorithm are available~\cite{HotSpot, G1GC}. The heap size is chosen to avoid getting ``Out of memory'' errors while running the benchmarks. The open file limit in Linux is increased to avoid getting ``Too many files open in the system'' error. The values of Spark internal parameters after tuning are given in Table~\ref{parameters}. Further details on the parameters are available~\cite{spark_config}.
  
\begin{table}[!ht]
\renewcommand{\arraystretch}{1.3}
\caption{JVM and Spark Parameters for Different Workloads.}
\label{parameters}
\centering
\begin{tabular}{p{1cm}|l|ccccc}
\multicolumn{2}{c|}{} & \multicolumn{1}{c|}{{\bf Wc}} & \multicolumn{1}{c|}{{\bf Gp}} & \multicolumn{1}{c|}{{\bf So}} & \multicolumn{1}{c|}{{\bf Km}} & {\bf Nb} \\ \hline
JVM & Heap Size (GB) & \multicolumn{5}{c}{50} \\ \cline{2-7} 
 & Old Generation Garbage Collector & \multicolumn{5}{c}{PS MarkSweep} \\ \cline{2-7} 
 & Young Generation Garbage Collector & \multicolumn{5}{c}{PS Scavange} \\ \hline
Spark & spark.storage.memoryFraction & \multicolumn{1}{c|}{0.1} & \multicolumn{1}{c|}{0.1} & \multicolumn{1}{c|}{0.1} & \multicolumn{1}{c|}{0.6} & 0.1 \\ \cline{2-7} 
 & spark.shuffle.memoryFraction & \multicolumn{1}{l|}{0.7} & \multicolumn{1}{l|}{0.7} & \multicolumn{1}{l|}{0.7} & \multicolumn{1}{l|}{0.4} & \multicolumn{1}{l}{0.7} \\ \cline{2-7} 
 & spark.shuffle.consolidateFiles & \multicolumn{5}{c}{true} \\ \cline{2-7} 
 & spark.shuffle.compress & \multicolumn{5}{c}{true} \\ \cline{2-7} 
 & spark.shuffle.spill & \multicolumn{5}{c}{true} \\ \cline{2-7} 
 & spark.shuffle.spill.compress & \multicolumn{5}{c}{true} \\ \cline{2-7} 
 & spark.rdd.compress & \multicolumn{5}{c}{true} \\ \cline{2-7} 
 & spark.broadcast.compress & \multicolumn{5}{c}{true} \\ \hline
\end{tabular}
\end{table}

\subsection{\textbf{Measurement Tools and Techniques}}

We configure Spark to collect GC logs which are then parsed to measure time (called real time in GC logs) spent in garbage collection. We rely on the log files generated by Spark to calculate the execution time of the benchmarks. We use Intel Vtune~\cite{Vtune} to perform concurrency analysis and general micro-architecture exploration. For scalability study, each benchmark is run 5 times within a single JVM invocation and the mean values are reported. For concurrency analysis, each benchmark is run 3 times within a single JVM invocation and Vtune measurements are recorded for the last iteration. This experiment is repeated 3 times and the best case in terms of execution time of the application is chosen. The same measurement technique is also applied in general architectural exploration, however the difference is that mean values are reported. Additionally, executor pool threads are bound to the cores before collecting hardware performance counter values. 

We use the top-down analysis method proposed by Yasin \cite{Top_Down_Method_for_Counters} to study the micro-architectural performance of the workloads. Super-scalar processors can be conceptually divided into the "front-end" where instructions are fetched and decoded into constituent operations, and the "back-end" where the required computation is performed. A pipeline slot represents the hardware resources needed to process one micro-operation. The top-down method assumes that for each CPU core, there are four pipeline slots available per clock cycle. At issue point each pipeline slot is classified into one of four base categories: Front-end Bound, Back-end Bound, Bad Speculation and Retiring. If a micro-operation is issued in a given cycle, it would eventually either get retired or cancelled. Thus it can be attributed to either Retiring or Bad Speculation respectively. Pipeline slots that could not be filled with micro-operations due to problems in the front-end are attributed to Front-end Bound category whereas pipeline slot where no micro-operations are delivered due to a lack of required resources for accepting more micro-operations in the back-end of the pipeline are identified as Back-end Bound.

\section{Scalability Analysis}

\subsection{Do Spark based data analytics benefit from using scale-up servers?}
We configure spark to run in local-mode and used system configuration parameters of Table~\ref{parameters}. Each benchmark is run with 1, 6, 12, 18 and 24 executor pool threads. The size of input data-set is 6 GB. For each run, we set the CPU affinity of the Spark process to emulate hardware with same number of cores as the worker threads. The cores are allocated from one socket first before switching to the second socket. Figure~\ref{speed-up} plots speed-up as a function of the number of cores. It shows that benchmarks scale linearly up to 4 cores within a socket. Beyond 4 cores, the workloads exhibit sub-linear speed-up, e.g., at 12 cores within a socket, average speed-up across workloads is 7.45. This average speed-up increases up to 8.74, when the Spark process is configured to use all 24 cores in the system. The performance gain of  mere 17.3\% over the 12 cores case suggest that Spark applications do not benefit significantly by using more than 12-core executors.

\subsection{Does performance remain consistent as we enlarge the data size?}

\begin{figure*}[!t]
\centering
\subfloat[Benchmarks do not benefit by adding more than 12 cores.]{\includegraphics[scale=0.29]{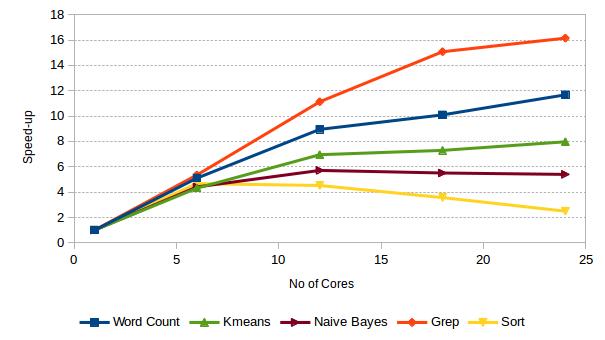}
\label{speed-up}}
\subfloat[Data processed per second decreases with increase in data size.]{\includegraphics[scale=0.29]{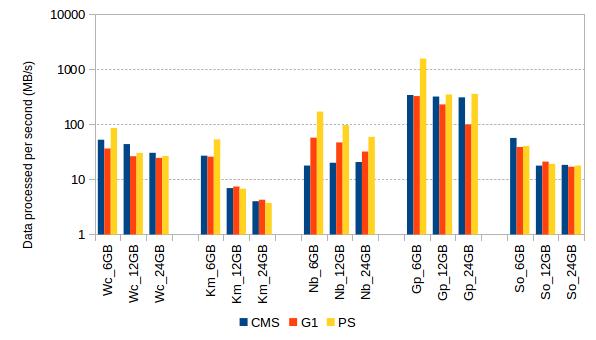}
\label{dps}}
\caption{Scale-up performance of applications: (a) when the number of cores increases and (b) when input data size increases.}
\label{scale-up performance}
\end{figure*}

The benchmarks are configured to use 24 executor pool threads in the experiment. Each workload is run with 6 GB, 12 GB and 24 GB of input data and the amount of data processed per second (DPS) is calculated by dividing the input data size by the total execution time. The data sizes are chosen to stress the whole system and evaluate the system's data processing capability. In this regard, DPS is a relevant metric as suggested in by Luo et al.~\cite{DPS}. We also evaluate the sensitivity of DPS to garbage collection schemes but explain it in the next section. Here we only analyse the numbers for Parallel Scavenge garbage collection scheme. By comparing 6 GB and 24 GB cases in Figure~\ref{dps}, we see that K-Means performs the worst as its DPS decreases by 92.94\% and Grep performs the best with a DPS decrease of 11.66\%. Furthermore, we observe that DPS decreases by 49.12\% on average across the workloads, when the data size is increased from 6 GB to 12 GB. However DPS decreases further by only 8.51\% as the data size is increased to 24GB. In the next section, we will explain the reason for poor data scaling behaviour.

\section{Limitations to Scale-up}

\subsection{How severe is the impact of garbage collection?}

\begin{figure*}[!t]
\centering
\subfloat[GC overhead is a scalability bottleneck.]{\includegraphics[scale=0.29]{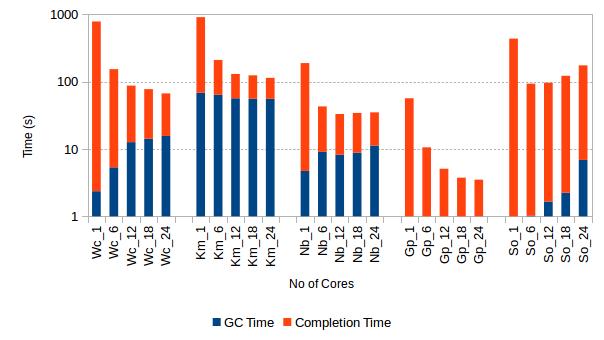}
\label{gc_6G}}
\subfloat[GC time increases at a higher rate with data size.]{\includegraphics[scale=0.29]{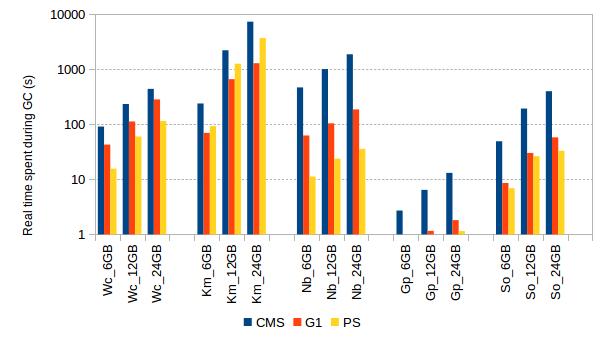}
\label{gc_datasize}}
\caption{Impact of garbage collection on application performance: (a) when the number of cores increases and (b) when input data size increases.}
\label{performance_gc}
\end{figure*}

Because of the in-memory nature of most Spark computations, garbage collection can become a bottleneck for Spark programs. To test this hypothesis, we analysed garbage collection time of scalability experiments from the previous section. Figure~\ref{gc_6G} plots total execution time and GC time across the number of cores. The proportion of GC time in the execution time increases with the number of cores. At 24 cores, it can be as high as 48\% in K-Means. Word Count and Naive Bayes also show a similar trend. This shows that if the GC time had at least not been increasing, the applications would have scaled better. Therefore we conclude that GC acts as a bottleneck.

To answer the question, ``How does GC affect data processing capability of the system?'', we examine the GC time of benchmarks running at 24 cores. The input data size is increased from 6 GB to 12 GB and then to 24 GB. By comparing 6 GB and 24 GB cases in Figure~\ref{gc_datasize}, we see that GC time does not increase linearly, e.g., when input data is increased by 4x, GC time in K-Means increases by 39.8x. A similar trend is also seen for Word Count and Naive Bayes. This also shows that if GC time had been increasing at most linearly, DPS would not have decreased significantly. For K-Means, DPS decreases by 14x when data size increases by 4x. For similar scenario in Naive Bayes, DPS decreases by 3x and GC time increases by 3x. Hence we can conclude that performance of Spark applications degrades significantly because GC time does not scale linearly with data size.

Finally we answer the question, ``Does the choice of Garbage Collector impact the data processing capability of the system?''. We look at impact of three garbage collectors on DPS of benchmarks at 6 GB, 12 GB and 24 GB of input data size. We study out-of-box (without tuning) behaviour of Concurrent Mark Sweep, G1 and Parallel Scavenge garbage collectors. Figure~\ref{gc_datasize} shows that across all the applications, GC time of Concurrent Mark Sweep is the highest and GC time of Parallel Scavenge is the lowest among the three choices. By comparing the DPS of benchmarks across different garbage collectors, we see that Parallel Scavenge results in 3.69x better performance than Concurrent Mark Sweep and 2.65x better than G1 on average across the workloads at 6 GB. At 24 GB, Parallel Scavenge performs 1.36x better compared to Concurrent Mark Sweep and 1.69x better compared to G1 on average across the workloads.

\subsection{Does file I/O become a bottleneck under large data volumes?}

\begin{figure*}[!t]
\centering
\subfloat[CPU utilization decreases with data size.]{\includegraphics[scale=0.29]{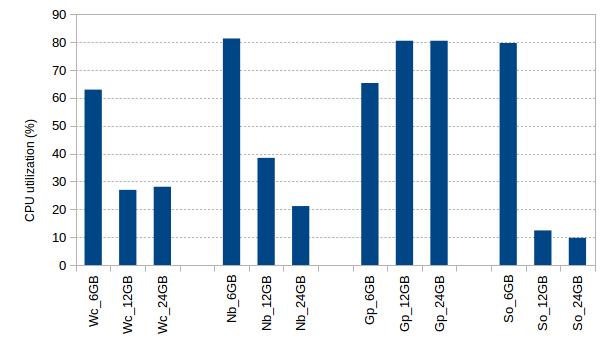}
\label{CPU_utilization}}
\subfloat[Wait time becomes dominant at larger datasets due to significant increase in file I/O operations.]{\includegraphics[scale=0.29]{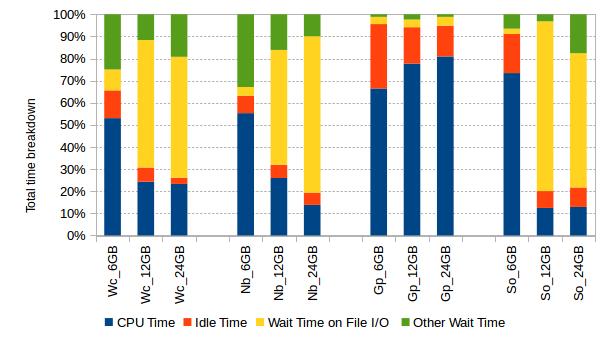}
\label{CPU_wait_time}}
\caption{Time breakdown under executor pool threads.}
\label{performance_thread_level}
\end{figure*}

In order to find the reasons for poor performance of Spark applications under larger data volumes, we studied the thread-level view of benchmarks by performing concurrency analysis in Intel Vtune. We analyse only executor pool threads as they contribute to 95\% of total CPU time during the entire run of the workloads. Figure~\ref{CPU_wait_time} shows that CPU time and wait time of all executor pool threads. CPU time is the time during which the CPU is actively executing the application on all cores. Wait time occurs when software threads are waiting on I/O operations or due to synchronization. The wait time is further divided into idle time and wait on file I/O operations. Both idle time and file I/O time are approximated from the top 5 waiting functions of executor pool threads. The remaining wait time comes under the category of ``other wait time''. 

It can be seen that the fraction of wait time increases with increase in input data size, except in Grep where it decreases. By comparing 6 GB and 24 GB case, the data shows that the fraction of CPU time decreases by 54.15\%, 74.98\% and 82.45\% in Word Count, Naive Bayes and Sort respectively; however it increases by 21.73\% in Grep. The breakdown of wait time reveals that contribution of file I/O increases by 5.8x, 17.5x and 25.4x for Word Count, Naive Bayes and Sort respectively but for Grep, it increases only 1.2x.   
The CPU time in Figure~\ref{CPU_wait_time} also correlates with CPU utilization numbers in Figure~\ref{CPU_utilization}. On average across the workloads, CPU utilization decreases from 72.34\% to 39.59\% as the data size is increased from 6 GB to 12 GB which decreases further by 5\% in 24 GB case.

\subsection{Is micro-architecture performance invariant to input data size?}

\begin{figure*}[!ht]
\centering
\subfloat[Retiring rate increases at larger datasets.]{\includegraphics[scale=0.29]{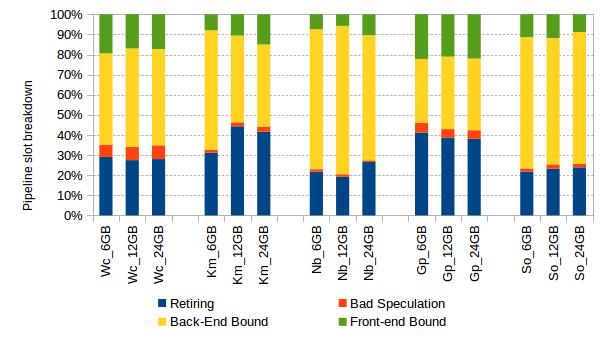}
\label{top_level}}
\subfloat[L1 Bound stalls increase with data size.]{\includegraphics[scale=0.29]{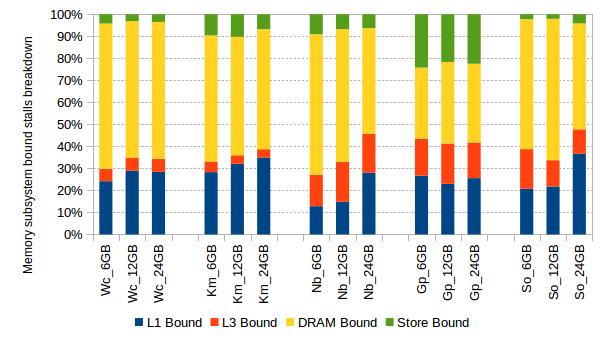}
\label{memory_level}}
\hfill
\subfloat[Port utilization increases at larger datasets.]{\includegraphics[scale=0.29]{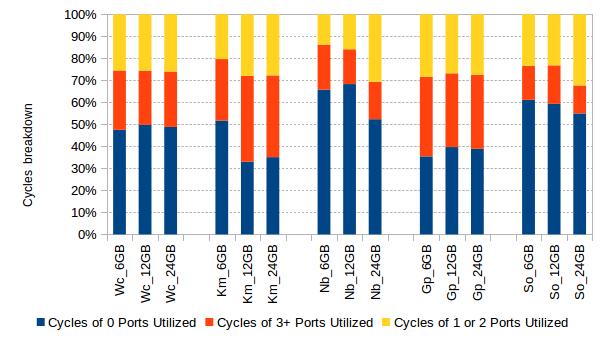}
\label{core_level}}
\subfloat[Memory traffic decreases with data size.]{\includegraphics[scale=0.29]{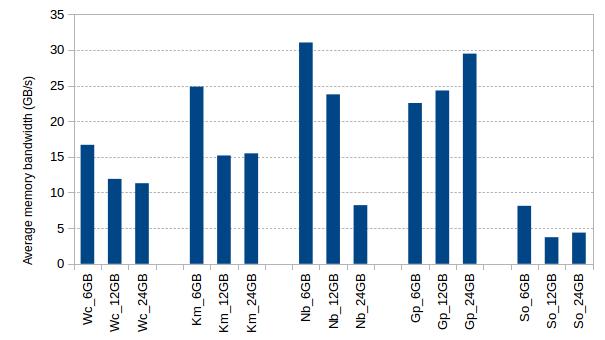}
\label{memory_bandwidth}}
\caption{Micro-architecture performance is inconsistent across different data sizes.}
\label{top_down_analysis}
\end{figure*}

We study the top-down breakdown of pipeline slots in the micro-architecture using the general exploration analysis in Vtune. The benchmarks are configured to use 24 executor pool threads. Each workload is run with 6 GB, 12 GB and 24 GB of input data. Figure~\ref{top_level} shows that benchmarks are back-end bound. On average across the workloads, retiring category accounts for 28.9\% of pipeline slots in 6 GB case and it increases to 31.64\% in the 24 GB case. Back-end bound fraction decreases from 54.2\% to 50.4\% on average across the workloads. K-Means sees the highest increase of 10\% in retiring fraction in 24 GB case in comparison to 6 GB case.

Next, we show the breakdown of memory bound stalls in Figure~\ref{memory_level}. The term DRAM Bound refers to how often the CPU was stalled waiting for data from main memory. L1 Bound shows how often the CPU was stalled without missing in the L1 data cache. L3 Bound shows how often the CPU was stalled waiting for the L3 cache, or contended with a sibling core. Store Bound shows how often the CPU was stalled on store operations. We see that DRAM bound stalls are the primary bottleneck which account for 55.7\% of memory bound stalls on average across the workloads in the 6 GB case. This fraction however decreases to 49.7\% in the 24 GB case. In contrast, the L1 bound fraction increase from 22.5\% in 6 GB case to 30.71\% in 24 GB case on average across the workloads. It means that due to better utilization of L1 cache, the number of simultaneous data read requests to the main memory controller decreases at larger volume of data. Figure~\ref{memory_bandwidth} shows that average memory bandwidth consumption decreases from 20.7 GB/s in the 6 GB case to 13.7 GB/s in the 24 GB case on average across the workloads.

Figure~\ref{core_level} shows the fraction of cycles during execution ports are used. Ports provide the interface between instruction issue stage and the various functional units. By comparing 6 GB and 24 GB cases, we observe that cycles during which no port is used decrease from 51.9\% to 45.8\% on average across the benchmarks and cycles during which 1 or 2 ports are utilized increase from 22.2\% to 28.7\% on average across the workloads.

\section{Related Work}

Several studies characterize the behaviour of big data workloads and identify the mismatch between the processor and the big data applications~\cite{Clearing_the_clouds,DCBench,BigDataBench,
understanding_in_memory_workloads, Characterising_and_subsetting_big_data_workloads, deep_dive_data_analytics,   MMU_Performance_Scale_out_workloads}. However these studies lack in identifying the limitations of modern scale-up servers for Spark based data analytics. Ousterhout et al.~\cite{Making_sense} have developed blocked time analysis to quantify performance bottlenecks in the Spark framework and have found out that CPU and not I/O operations are often the bottleneck. Our thread level analysis of executor pool threads shows that the conclusion made by Ousterhout et al. is only valid when the the input data-set fits in each node's memory in a scale-out setup. When the size of data set on each node is scaled-up, file I/O becomes the bottleneck again. Wang et al.~\cite{BigDataBench} have shown that the volume of input data has considerable affect on the micro-architecture behaviour of Hadoop based workloads. We make  similar observation about Spark based data analysis workloads.

\section{Conclusions}

We have reported a deep dive analysis of Spark based data analytics on a large scale-up server.
The key insights we have found are as follows:

\begin{itemize}
\item Spark workloads do not benefit significantly from executors with more than 12 cores. 
\item The performance of Spark workloads degrades with large volumes of data due to substantial increase in garbage collection and file I/O time.
\item With out any tuning, Parallel Scavenge garbage collection scheme outperforms Concurrent Mark Sweep and G1 garbage collectors for Spark workloads.
\item Spark workloads exhibit improved instruction retirement due to lower L1 cache misses and better utilization of functional units inside cores at large volumes of data.
\item Memory bandwidth utilization of Spark benchmarks decreases with large volumes of data and is 3x lower than the available off-chip bandwidth on our test machine.
\end{itemize}
 
We conclude that Spark run-time needs node-level optimizations to maximize its potential on modern servers. Garbage collection is detrimental to performance of in-memory big data systems and its impact could be reduced by careful matching of garbage collection scheme to workload. Inconsistencies in micro-architecture performance across the data sizes pose additional challenges for computer architects. Off-chip memory buses should be optimized for in-memory data analytics workloads by scaling back unnecessary bandwidth.

\bibliographystyle{splncs03}
\bibliography{references}

\end{document}